\PassOptionsToPackage{dvipsnames}{xcolor} 
\RequirePackage{etoolbox}
\documentclass[sigconf, 9pt, authorversion, natbib=false, language=british, pbalance=true]{acmart} 
\usepackage[inline]{enumitem}
\usepackage{multirow}
\usepackage[fixed]{fontawesome6}
\usepackage{tikz}
\usetikzlibrary{arrows.meta, backgrounds, calc, positioning}
\usepackage{minted} 

\usepackage{twemojis}

\usepackage{subcaption}
\usepackage[per-mode=symbol]{siunitx}
\DeclareSIUnit{\Wh}{Wh}
\DeclareSIUnit{\mA}{mA}
\DeclareSIUnit{\mAh}{mAh}
\DeclareSIUnit\year{yr}
\DeclareSIUnit{\tCOeq}{tCO_2e}
\DeclareSIUnit{\kgCOeq}{kgCO_2e}
\DeclareSIUnit{\kWh}{kWh}
\DeclareSIUnit{\gCOeq}{gCO_2e}
\DeclareSIUnit{\NOx}{NO_x}
\DeclareSIUnit{\pkt}{pkt}
\usepackage{textcomp}

\RequirePackage[datamodel=acmdatamodel, style=acmnumeric, sortcites]{biblatex}
\usepackage[acronym]{glossaries} 
\glsdisablehyper 
\newacronym{co2e}{\(\mathrm{CO_{2}e}\)}{\(\mathrm{CO_{2}}\)-equivalent}
\newacronym{ecc}{ECC}{Error Correcting Code}
\newacronym{ghg}{GHG}{greenhouse gas}
\newacronym{gsl}{GSL}{ground-satellite link}
\newacronym{isl}{ISL}{inter-satellite link}
\newacronym{lcsa}{LCSA}{Life Cycle Sustainability Assessment}
\newacronym{leo}{LEO}{low-Earth Orbit}
\newacronym{mios}{MIOS}{Moon Ice Observation Satellite}
\newacronym{sci}{SCI}{Software Carbon Intensity}
\newacronym{sdg}{SDG}{Sustainable Development Goal}

\definecolor{tol-hc-yellow}{RGB}{221, 170, 51} 
\definecolor{tol-hc-red}{RGB}{187, 85, 102} 
\definecolor{tol-hc-blue}{RGB}{0, 68, 136} 
\definecolor{c-gas}{RGB}{135, 206, 250}
\definecolor{c-oil}{RGB}{77, 77, 77}
\definecolor{todo-galapagos}{RGB}{153, 214, 76}

\usepackage{soul}
\newcommand{\sThree}{\texttt{ESpaS}}

\usepackage[most]{tcolorbox}
\newcommand{\insightbox}[1]{
  \begin{tcolorbox}[title=,title filled=false,colback=todo-galapagos!15!white,colframe=todo-galapagos!75!black,arc=.3em,boxsep=-1mm]
    \parbox{\columnwidth}{%
        \textbf{Insight:} {#1}
    }
\end{tcolorbox}
}

\usepackage{todonotes} 

\newfloat{mintedfloat}{htbp}{lop}
\floatname{mintedfloat}{Listing}

\setcopyright{cc}
\setcctype[4.0]{by}
\copyrightyear{2025}
\acmYear{2025}
\acmDOI{10.1145/3757892.3757896}

\acmConference[ACM SIGENERGY Energy Inform. Rev.]{ACM SIGENERGY Energy Informatics Review}{Volume 5 Issue 2}{July 2025}
\acmBooktitle{ACM SIGENERGY Energy Informatics Review}
\acmISBN{}

\begin{document}
\renewcommand{\sectionautorefname}{Section} 
\renewcommand{\subsectionautorefname}{Section} 
\renewcommand{\equationautorefname}{Eq.} 
\setlength{\marginparwidth}{15mm} 

\title[Dirty Bits in Low-Earth Orbit]{Dirty Bits in Low-Earth Orbit:\texorpdfstring{\\}{ }The Carbon Footprint of Launching Computers}

\author{Robin Ohs}
\orcid{0009-0002-1436-6196}
\author{Gregory F. Stock}
\orcid{0000-0001-5170-2019}
\author{Andreas Schmidt}
\orcid{0000-0002-7113-7376}
\affiliation{%
    \institution{Saarland University}
    \city{Saarbrücken}
    \country{Germany}}

\author{Juan A. Fraire}
\orcid{0000-0001-9816-6989}
\affiliation{%
    \institution{Inria Lyon}
    \city{Villeurbanne}
    \country{France}}
\affiliation{%
    \institution{Saarland University}
    \city{Saarbrücken}
    \country{Germany}}

\author{Holger Hermanns}
\orcid{0000-0002-2766-9615}
\affiliation{%
    \institution{Saarland University}
    \city{Saarbrücken}
    \country{Germany}}

\renewcommand{\shortauthors}{Ohs et al.}

\begin{abstract}
    \Gls{leo} satellites are increasingly proposed for communication and in-orbit computing, achieving low-latency global services.
However, their sustainability remains largely unexamined.
This paper investigates the carbon footprint of computing in space, focusing on lifecycle emissions from launch over orbital operation to re-entry.
We present \sThree{}, a lightweight tool for estimating carbon intensities across CPU usage, memory, and networking in orbital vs. terrestrial settings.
Three worked examples compare
\begin{enumerate*}[label=(\roman*)]
    \item launch technologies (state-of-the-art rocket vs.\@ potential next generation),
    \item operational emissions of data center workloads in orbit and on the ground and,
    \item in-orbit aggregation with raw data transmission.
\end{enumerate*}
Results show that, even under optimistic assumptions, in-orbit systems incur significantly higher carbon costs---primarily due to embodied emissions from launch and re-entry.
Our findings advocate for carbon-aware design principles and regulatory oversight in developing sustainable digital infrastructure in orbit.

\glsresetall 

\end{abstract}


\settopmatter{printacmref=false, printccs=true, printfolios=true}

\maketitle


\section{Introduction}

We are witnessing a new space race~\cite{bhattacherjee2018gearing}, driven not (only) by geopolitical rivalry but by the convergence of commercial ambition and innovation in networking and emerging space-based services.
In recent years, large-scale \gls{leo} constellations have been rapidly deployed to deliver global Internet access.
These systems promise transformative benefits: worldwide geographic coverage~\cite{li2021internet}, ultra-low latency connectivity for interactive applications~\cite{handley2018delay}, and the emergence of vertically integrated satellite ISPs capable of rivaling terrestrial operators~\cite{klenze2018networking}.

This shift has not occurred in isolation.
On the ground, hybrid and distributed ground station networks are being rearchitected to meet the demands of continuous \gls{leo} tracking and data offloading~\cite{vasisht2020distributed}.
In parallel, architectural innovations such as using terrestrial relays to optimize end-to-end latency~\cite{handley2019using} and exploiting \glspl{isl} to build in-orbit backbones~\cite{hauri2020internet} reveal the tight coupling between space and terrestrial infrastructure.

However, as \gls{leo} networks scale, fundamental limits in physical space and link interference are beginning to surface~\cite{chen2024unraveling}, demanding new network design, coordination, and sustainability paradigms.
The risk is not only technical but systemic.
As highlighted by \citeauthor*{muriga2024road}~\cite{muriga2024road}, the current trajectory of \gls{leo} development is consolidating power in the hands of a few dominant players.
These vertically integrated actors—often controlling orbital infrastructure and end-user services—pose risks to openness, interoperability, and fair access.
In response, some have advocated for decentralized models of satellite operation and ownership~\cite{oh2024call}; yet, such approaches may inadvertently accelerate the deployment of redundant or poorly coordinated constellations.
A key concern is the escalating accumulation of space debris~\cite{pardini2020environmental, esa:spaceEnvReport25}, but mounting warnings also point to radio astronomy interference~\cite{abedi2024foe} and the environmental toll of satellite manufacturing~\cite{kumaran2024quantifying} as critical sustainability threats.
Without cohesive governance and shared architectural principles, \gls{leo} networks face a growing tension between scalability and sustainability.

While governance, orbital congestion, and electromagnetic interference are well-recognized challenges in the discourse on \gls{leo} constellations, a critical dimension remains conspicuously underexplored: the carbon footprint of orbital infrastructure.
Recent evidence by \citeauthor*{osoro2023sustainability}~\cite{osoro2023sustainability} shows that \gls{leo} satellite broadband systems exhibit lifecycle emissions per subscriber of up to 14 times the footprint of terrestrial mobile broadband.
This challenges prevailing assumptions about the environmental neutrality of satellite-based digital infrastructure, especially as mega-constellations scale to tens of thousands of satellites~\cite{gaston2023environmental}.
Furthermore, \citeauthor*{marc2024comprehensive}~\cite{marc2024comprehensive} show that, in several space and Earth science institutions, satellite operations account for the largest share of institutional emissions—surpassing those from air travel, daily commuting, and laboratory activities.
These findings expose a fundamental research and policy gap: carbon emissions from space activities are rarely quantified, reported, or incorporated into sustainability frameworks.

Addressing this blind spot is the central objective of this work.
We argue that the carbon footprint of space infrastructure must be treated as a first-class sustainability concern—on par with orbital debris and spectral congestion—particularly as space systems become increasingly entangled with terrestrial digital infrastructure.
To anticipate the sustainability implications of this trend, we present a series of back-of-the-envelope analyses quantifying the carbon intensity of computing in space.
This approach is both necessary and deliberate: detailed lifecycle and operational data for orbital systems are often difficult to obtain—frequently more so than for terrestrial infrastructure—due to proprietary constraints, limited disclosure, and fragmented reporting practices.

We therefore focus on three worked examples:
\begin{enumerate}
    \item \textbf{Launch Technologies}: We compare the lifecycle emissions of conventional rocket-based launches~(e.g. Falcon-9) with hypothetical alternatives that propose, e.g., higher payload to rocket mass ratio to lower environmental costs.
    \item \textbf{Data Center Workloads in Orbit}: We assess the carbon footprint of data centers workloads in orbit, estimating\linebreak emission intensities across three key operational dimensions: compute time (CPU: \unit{\gCOeq\per\second}), storage~(DRAM/SSD: \unit{\gCOeq\per\giga\byte\per\second}), and networking~(NIC: \unit{\gCOeq\per\pkt}).
    \item \textbf{In-Orbit Data Aggregation}: We compare the in-orbit processing and aggregation of data with downloading unaggregated raw data and processing it on Earth---highlighting the trade-off between networking and computing.
\end{enumerate}
These metrics will expose the implicit carbon footprint of moving computing to orbit, compared to optimized terrestrial equivalents.

To support this analysis, we introduce \sThree{}~(Estimator for Space Sustainability), a lightweight tool to estimate the energy and carbon of orbital computing under varying assumptions.
Preliminary evaluations reveal key trade-offs in system design, launch strategy, and workload placement.
Together, these results aim to lay the groundwork for carbon-aware design principles and regulatory frameworks in the next generation of satellite computing ecosystems.

In the remainder of the paper, \autoref{sec:background} gives further information on sustainability, carbon emissions in space as well as launching computers to orbit and operating them there.
Our models and estimation tool \sThree{} are described in \autoref{sec:models} and applied to three worked examples in \autoref{sec:worked-examples}.
This section reveals several key insights, which are further reflected upon in the concluding \autoref{sec:conclusion}.

\section{Background}
\label{sec:background}

The growing attention to sustainability in space aligns with global policy frameworks such as the United Nations \glspl{sdg}~\cite{un:sdgs} and the European Union's Green Deal~\cite{eu:green-deal}---both emphasize climate action and responsible innovation.
The carbon footprint (measured in mass of \acrlongpl{co2e}\glsunset{co2e}~(e.g.\@ \unit{\kgCOeq})) has become a central metric for assessing the environmental impact of space systems. 
It quantifies the total \gls{ghg} emissions, normalized to the warming potential of \(\mathrm{CO_{2}}\), associated with the lifecycle of a product or service.

\subsection{Carbon Emissions in Space Missions}

A recent quantitative study by \citeauthor*{osoro2023sustainability}~\cite{osoro2023sustainability} assesses the environmental sustainability of \gls{leo} broadband mega-con\-stel\-la\-tions, providing one of the first lifecycle carbon estimates per user.
The authors model the emissions associated with satellite production, launch, and operation for systems like Starlink, estimating a worst-case footprint of up to \qty{469}{\kgCOeq} per subscriber per year.
Even under average assumptions, the carbon intensity reaches \qty{250}{\kilo\gCOeq} per subscriber annually---roughly 6 to 8 times higher than the terrestrial benchmark of \qtyrange{32.8}{39.5}{\kgCOeq}.
Their methodology integrates network-level simulations with lifecycle analysis, revealing that satellite-based connectivity can expand global access.
However, it also imposes substantial carbon externalities that challenge the perception of \gls{leo} systems as a sustainable digital infrastructure.

A recent comprehensive study by \citeauthor*{marc2024comprehensive}~\cite{marc2024comprehensive} highlights that satellite infrastructures represent a dominant and previously underappreciated contributor to the carbon footprint of Earth and space science laboratories.
Through an analysis of six French research institutions, the study reveals that satellite usage alone can account for up to 65\% of a laboratory's total annual emissions, with per-capita values reaching \qty{12}{\tCOeq}—well above the \qty{2}{\tCOeq} per year target aligned with limiting global warming to \qty{1.5}{\degreeCelsius}.
These emissions surpass those from travel, computing, and building operations.
By introducing a methodology to allocate emissions from shared infrastructures, such as satellite missions, to individual research units based on publication output, the authors demonstrate that current sustainability assessments in science vastly underestimate the role of space-based assets.
This underscores the urgent need to integrate satellite-related emissions into sustainability metrics and policy frameworks, especially as space infrastructure becomes increasingly central to scientific, commercial, and societal functions.

\citeauthor*{gaston2023environmental}~\cite{gaston2023environmental} present a comprehensive lifecycle assessment---extending well beyond orbital congestion.
Their analysis spans from resource extraction and manufacturing to launch emissions, orbital operations, and decommissioning, highlighting key concerns such as aluminum-based atmospheric pollution during satellite burn-up, climate and ozone impacts from rocket exhaust, and the long-term ecological implications of night-sky brightness.
The study also identifies the embedded carbon and material costs of ground-based infrastructure supporting satellite services, including energy-intensive cloud computing and data distribution systems.

\citeauthor*{wilson2023implementing}~\cite{wilson2023implementing} introduce a comprehensive \gls{lcsa} framework for space missions.
Their study also highlights the critical need to quantify carbon emissions early in the mission design process.
The authors explicitly incorporate \gls{ghg} emissions—expressed in \gls{co2e}—within their sustainability scoring metrics and use carbon footprinting as a core environmental indicator.
In their case study of the \gls{mios}, they estimate mission-wide \gls{co2e} emissions across different design configurations and use these estimates to inform sustainability trade-offs.
Although the study does not perform a high-resolution carbon accounting per subsystem, it exemplifies how early integration of carbon metrics in decision-making can significantly shape the sustainability profile of space missions.

\subsection{Data Centers in Orbit}

Actual research~\cite{bhattacherjee2020orbit,ascend2025,axdcu2025} explores the idea of using large \gls{leo} satellite constellations as platforms for in-orbit computing, effectively transforming and extending communication satellites into distributed, cloud-like compute nodes.
Their thought experiment considers a spectrum of use cases, from edge content delivery~(see also~\cite{bose2024s}) and multiplayer gaming to onboard preprocessing of space-native data.
While power constraints, heat dissipation, and satellite lifecycle impose nontrivial challenges, the authors argue that neither mass, volume, nor radiation hardening pose fundamental barriers to feasibility.
They demonstrate that leveraging otherwise idle \gls{leo} assets for compute tasks can improve constellation utility and reduce bandwidth demand for downlinking voluminous raw sensing data—contributing, indirectly, to a more efficient and potentially sustainable use of orbital infrastructure.

\section{Modeling \& Estimating Sustainability}
\label{sec:models}

To model the sustainability of space systems with respect to carbon emissions, we follow the \emph{\gls{sci}} methodology~\cite{sci}.
For workload $R$~(of arbitrary, but fixed unit), carbon emissions $C$ (in \unit{\gCOeq}) are emitted: an intensity of $I = \frac{C}{R}$.
$C$ is the sum of operational~($O$) and embodied~($M$) emissions, hence $I = \frac{O + M}{R}$ or $I = I_O + I_M$ with $R$ factored into the individual components.
In detail, $M = \mathit{TE} \cdot \mathit{RS} \cdot \mathit{TS}$, i.e. the emissions~$M$ for an $R$ are the total emissions~($\mathit{TE}$) multiplied with a resource share~($\mathit{RS}$), if the resource is divisible, and a time share~($\mathit{TS}$).
The latter is important, as space missions are often strictly time-bound (unlike terrestrial applications).
When searching for realistic $\mathit{TE}$ values, we often find data sheets that include production, transport, and disposal emissions.
However, for orbital systems, the numbers found in these data sheets are too small, as launch technologies significantly increase embodied emissions~(precisely: the transport portion).
We will now model and estimate these aspects.

\subsection{Modeling}

\mathchardef\mhyphen="2D 

\newcommand{\vMission}{\ensuremath{\mathit{Mission}}}
\newcommand{\vProp}{\ensuremath{\mathit{Prop}}}
\newcommand{\vPayload}{\ensuremath{\mathit{Payl}}}
\newcommand{\vLaunch}{\ensuremath{\mathit{Lch}}}
\newcommand{\vReentry}{\ensuremath{\mathit{Re\mhyphen{}Entry}}}
\newcommand{\vFirst}{\ensuremath{\mathit{1st}}}
\newcommand{\vSecond}{\ensuremath{\mathit{2nd}}}
\newcommand{\vEarth}{\ensuremath{\mathit{Earth}}}
\newcommand{\vRocket}{\ensuremath{\mathit{F9}}}
\newcommand{\vStarshipN}{\ensuremath{\mathit{StShN}}}
\newcommand{\vStarship}{\ensuremath{\mathit{StSh}}}
\newcommand{\vSpin}{\ensuremath{\mathit{Spin}}}
\newcommand{\vEnergy}{\ensuremath{\mathit{Energy}}}
\newcommand{\vHydrogen}{\ensuremath{\mathit{Hydrogen}}}
\newcommand{\vSolarArray}{\ensuremath{\mathit{Solar}}}
\newcommand{\vProduction}{\ensuremath{\mathit{Prod}}}
\newcommand{\vBattery}{\ensuremath{\mathit{Bat}}}
\newcommand{\vLifetime}{\ensuremath{\mathit{Life}}}
\newcommand{\vCompute}{\ensuremath{\mathit{Comp}}}
\newcommand{\vCPU}{\ensuremath{\mathit{CPU}}}
\newcommand{\vMemory}{\ensuremath{\mathit{Mem}}}
\newcommand{\vDRAM}{\ensuremath{\mathit{DRAM}}}
\newcommand{\vPacket}{\ensuremath{\mathit{Pkt}}}
\newcommand{\vDataRate}{\ensuremath{\mathit{DataRate}}}
\newcommand{\vTransmission}{\ensuremath{\mathit{Transmission}}}
\newcommand{\vTransceiver}{\ensuremath{\mathit{Transceiver}}}

We assume that a system is using a launch technology~$t$ and is dimensioned for a mission time $T_{\vMission}$.

\subsubsection*{Launch \& Re-Entry Emissions}\hfill

First, we estimate $I_{\vLaunch, t}$ in \unit{\kgCOeq\per\kilo\gram} for different launch technologies~$t$.
For a component~$c$ with mass~$m_{\mathit{c}}$, the launch-related embodied carbon will be $\mathit{TE}_{\vLaunch, c, t} = m_{c} \cdot I_{\vLaunch, t}$ in \unit{\kgCOeq} (if $t$ is omitted for brevity, we assume the current system's technology).
Assuming a two-stage ($\vFirst$, $\vSecond$) rocket with reusability factors $n_1$, $n_2$, we get:
\begin{align}
    C_{\vLaunch, t}(n_1, n_2) & = \frac{\mathit{TE}_{\vProduction, \vFirst, t}}{n_1} + \frac{\mathit{TE}_{\vProduction, \vSecond, t}}{n_2} + \mathit{TE}_{\vProp, t} \nonumber \\
    I_{\vLaunch, t}(n_1, n_2) & = \frac{C_{\vLaunch, t}(n_1, n_2)}{m_{\vPayload, t}} \label{eq:intensity-launch}
\end{align}

Re-entry ablation of discarded objects (e.g. spacecraft and upper stages) produces among others alumina particles (\(\mathrm{Al_{2}O_{3}}\)) and gaseous reactive nitrogen (\unit{\NOx})~\cite{journals/scidata/BarkerMM24, journals/actaastro/ParkNLM21}.
Although \(\mathrm{Al_{2}O_{3}}\) does contribute to global warming---primarily as it damages the Ozone layer~\cite{journals/grl/Ferreira24}---we currently solely consider \unit{\NOx} in our calculations.
The warming effect caused by \(\mathrm{Al_{2}O_{3}}\) is of a different nature, making its global warming potential hard to express in \gls{co2e}.
To avoid exaggerating the effect, we currently leave it out numerically, but would incorporate it, once there is a conversion to \gls{co2e}.

The amount of \unit{\NOx} emitted during re-entry is proportional to the total mass of the re-entering object (and also depends on the re-entry trajectory and velocity)~\cite{journals/scidata/BarkerMM24}.
We get the following general intensity per re-entering non-reusable mass:
\[
    I_{\vReentry} = \qty{0.4}{\kg\NOx\per\kg} = \qty{13.2}{\kgCOeq\per\kg}
\]
For a concrete launch technology~\(t\), we would normalize with respect to the payload, where $m_{\vSecond, t}$ depends on the launch technology, e.g. (classical) second stage for Falcon-9.
\[
    I_{\vReentry, t} = \frac{m_{\vPayload} + m_{\vSecond, t}}{m_{\vPayload, t}} \cdot I_{\vReentry}
\]

\paragraph{\textcolor{tol-hc-red}{\faIcon{rocket} Falcon-9 (F9)}}

The first and second stage of a Falcon 9 v1.2 use a total of \qty{155,87}{\tonne} of RP-1 fuel~\cite{spaceflight101:Falcon9FT}.
RP-1 emits $\qty[per-mode=symbol]{3.143}{\kgCOeq\per\kg}$ during combustion under idealized conditions~\cite{journals/jtht/Wang01}, resulting in total, propellant-induced emissions of \(\mathit{TE}_{\vProp, \vRocket} = \qty{489.9}{\tCOeq}\).
The raw materials of the reusable first stage account for $\mathit{TE}_{\vProduction, \vFirst, \vRocket} = \qty{423.5}{\tCOeq}$ and the non-reusable second stage for $\mathit{TE}_{\vProduction, \vSecond, \vRocket} = \qty{109.7}{\tCOeq}$.
Assuming a (publicly confirmed~\cite{web/twitter/falcon9-payload}) $m_{\vPayload,\vRocket} = \qty{17.5}{\tonne}$ and $n_1 = 20$ starts with a reusable first stage, we get:
\[
    I_{\vLaunch, \vRocket} = \frac{C_{\vLaunch, \vRocket}(20, 1)}{m_{\vPayload, \vRocket}} \approx\qty{35.5}{\kgCOeq\per\kilo\gram} 
\]
For re-entry, we consider the second stage and the spacecraft that are burned in the atmosphere (the payload fairing lands in the ocean and can be recovered):
\[
    I_{\vReentry, \vRocket} = \frac{\qty{17.5}{\tonne} + \qty{4.0}{\tonne}}{\qty{17.5}{\tonne}} \cdot I_{\vReentry} \approx \qty{16.2}{\kgCOeq\per\kg}
\]

Hence, the overall Falcon-9 launch and re-entry emissions are:
\[
    I_{L\&R, \vRocket} = I_{\vLaunch, \vRocket} + I_{\vReentry, \vRocket} \approx \qty{51.7}{\kgCOeq\per\kg}
\]

Though numbers are specific for Falcon-9, we have this insight:
\insightbox{Re-entry is often overlooked, but adds further emissions on top---about another half of the launch emissions.}

\paragraph{\textcolor{tol-hc-blue}{\faIcon{rocket} Starship (StSh)}}
For some time now, SpaceX has been developing Starship, a ``two-stage \emph{fully reusable} super heavy-lift launch vehicle''~\cite{web/wikipedia/starship, web/spacex/starship, web/faa/starship-eval25-final}.
With a second stage mass of $m_{\vSecond,\vStarship} = \qty{85.5}{\tonne}$ and a best-case payload of $m_{\vPayload,\vStarship} = \qty{150}{\tonne}$ the proportions change.
Lacking data on production emissions, we linearly extrapolate the first and second stage embodied emissions based on their mass increase over Falcon-9 ($\vFirst \times 12.4$ and $\vSecond \times 14.8$).
The amount of propellant increases as well to, reportedly, \qty{3870}{\tonne} of liquid oxygen~($O_2$) and \qty{1030}{\tonne} of liquid methane~($\mathit{CH}_4$).
This is a different mix from \mbox{Falcon-9} ($\mathit{CH}_4$ vs.\@ RP-1 \cite{web/spacex/starship,web/spacex/falcon-9}).
Combusting liquid methane has an intensity of \qty{2.75}{\kgCOeq\per\kg}, hence:
\[
    \mathit{TE}_{\vProp,\vStarship} = \qty{1030}{\tonne} \cdot \qty{2.75}{\kgCOeq\per\kg} = \qty{2832.5}{\tCOeq}
\]
This gives us (normalized by payload) the following intensity:
\[
    I_{\vProp,\vStarship} = \qty{18.9}{\kgCOeq\per\kilo\gram}
\]
With $\mathit{TE}_{\vProduction,\vFirst,\vStarship} = \qty{5245}{\tCOeq}$, $\mathit{TE}_{\vProduction,\vSecond,\vStarship} = \qty{1621}{\tCOeq}$, and reuse $n_1 = n_2 = 20$, \autoref{eq:intensity-launch} gives us $I_{\vLaunch, \vStarship} \approx \qty{21.2}{\kgCOeq\per\kilo\gram}$.

Due to full reusability of everything that is not the payload, we get the optimal re-entry intensity
$I_{\vReentry, \vStarship} = I_{\vReentry}$
and hence overall intensity:
\[
    I_{L\&R, \vStarship} = I_{\vLaunch, \vStarship} + I_{\vReentry, \vStarship} \approx\qty{34.4}{\kgCOeq\per\kg}
\]


\subsubsection*{Power Supply's Energy Intensity}\hfill

With carbon emissions mainly associated with the cleanness of energy used in processes, it is evident that the system's power supply is a major contributor to overall emissions.

A satellite is typically equipped with a \emph{Solar Array} and a \emph{Battery}---note that we employ the same battery on Earth to cover clouds and night times.
For the rest of the system, an energy carbon intensity $I_{\vEnergy}$ in \unit{\gCOeq\per\kWh} is the sum of its two parts $I_{\vEnergy, \vBattery}$ and $I_{\vEnergy, \vSolarArray}$.

For the battery, the total lifetime energy to be (dis-)charged from/to it is $E_{\vLifetime, \vBattery} = \mathit{Cycles} \cdot \mathit{Capacity}$ (here, the time share is in battery lifetime, not absolute time).
The total embodied emissions are $\mathit{TE}_{\vBattery,t} = \mathit{TE}_{\vProduction, \vBattery} + \mathit{TE}_{L\&R, t}(m_{\vBattery})$ for a launch technology $t$ and battery mass $m_{\vBattery}$.
While concrete values vary, we approximate an energy density of \qty{3.75}{\kilo\gram\per\kWh}~\cite{web/thunder-said-energy/energy-density,hasan2025} and battery cradle-to-gate emissions of $\mathit{TE}_{\vProduction, \vBattery}(\mathit{cap}) = \mathit{cap} \cdot \qty{100}{\kgCOeq\per\kWh}$~\cite{kallitsis2024-li-battery,peiseler2024}.
Given a capacity, cycles, and launch technology, we calculate $m_{\vBattery}$ and arrive at an intensity of $I_{\vEnergy, \vBattery, t} = \frac{\mathit{TE}_{\vBattery,t}}{E_{\vLifetime, \vBattery}}$.

For the solar array, with a similar computation using~\cite{web/sparkwing/solar}, we compute as $m_{\vSolarArray} = n_{\mathit{panels}} \cdot (\qty{3.41}{\kilo\gram\per\square\meter} + \qty{0.65}{\kilo\gram}) + \qty{0.4}{\kilo\gram}$~(fixed cost for central control and per panel, together with an area-proportional part).
Using an online source~\cite{web/etude/solar}, we get $\mathit{TE}_{\vProduction,\vSolarArray}(\mathit{power}) = \mathit{power} \cdot \qty{615}{\kgCOeq\per\kilo\watt}$.
In consequence, we get total embodied emissions $\mathit{TE}_{\vSolarArray, t} = \mathit{TE}_{\vProduction, \vSolarArray} + \mathit{TE}_{L\&R, t}(m_{\vSolarArray})$ for a launch technology $t$.
For the provided power, we assume per panel to have $P_{\vSolarArray, \mathit{Earth}} = \qty{400}{\watt\per\square\meter}$ and $P_{\vSolarArray, \mathit{Orbit}}= \qty{1367}{\watt\per\square\meter}$ due to a thinner atmosphere in orbit (location omitted in the following for brevity, if unambiguous).
Given a mission time, we can supply a maximum $E_{\vLifetime, \vSolarArray} = T_{\vMission} \cdot P_{\vSolarArray}$.
Hence, again, $I_{\vEnergy, \vSolarArray, t} = \frac{\mathit{TE}_{\vSolarArray, t}}{E_{\vLifetime, \vSolarArray}}$.

\subsubsection*{Compute Intensity}
A computing system uses energy to execute computations.
In space, this is mostly done with CPUs and FPGAs~\cite{web/amd/versal-asoc}.
Given an average power $P_{\mathit{CPU}}$ and a die area $A_{\vCPU}$, we can derive $I_{\vCompute}$ in \unit{\gCOeq\per\second} of CPU time.
Operational emissions are relatively straightforward $\mathit{OI}_{\mathit{CPI}} = P_{\vCPU} \cdot I_{\vEnergy}$.
Note that we assume that using it a certain time means using all cores to maximum extend.
If a smaller granularity is needed, core time should be used instead as a base unit.
On the embodied side, it is necessary to quantify $\mathit{TE}_{\vProduction, \vCPU}$ and mass $m_{\vCPU}$.
For the former, we consider ACT~\cite{DBLP:conf/isca/GuptaEHWL0W22}, where we find an approximation of $\mathit{CPA} = \qty{250}{\gCOeq\per\square\centi\meter}$, hence $\mathit{TE}_{\vProduction, \vCPU} = \mathit{CPA} \cdot A_{\vCPU}$.
In consequence, $\mathit{MI}_{\vCPU,t} = \frac{\mathit{TE}_{\vProduction, \vCPU} + \mathit{TE}_{L\&R,t}(m_{\vCPU})}{T_{\vMission}}$.

\subsubsection*{Store Intensity}
A computing system uses energy also to store intermediate results (in DRAM) or long-term results (on ``disks'' or comparable media).
DRAM usage is measured in \emph{held memory} $(V,T)$ in (\unit{\byte}, \unit{\second}), hence $I_{\vDRAM}$ in \unit{\gCOeq\per\byte\per\second}.
The DRAM is configured with a data volume $V_{\vDRAM}$ (capacity) in \unit{\giga\byte} and a power per held memory $P_{\vDRAM}/V_{\vDRAM}$, e.g. in \unit{\milli\watt\per\giga\byte}.
The intensities are:
\[
    \mathit{OI}_{\vDRAM} = \frac{P_{\vDRAM}}{V_{\vDRAM}} \cdot I_{\vEnergy} \;;\; \mathit{MI}_{\vDRAM, t} = \frac{\mathit{TE}_{\vProduction, \vDRAM} + \mathit{TE}_{L\&R,t}}{T_{\vMission} \cdot V_{\vDRAM}}
\]
Further, memory in space is error-prone, so it is deployed redundantly or employs error correction~(e.g. \gls{ecc} memory).
Both approaches add overhead in terms of production or operation, inducing higher emissions and lower efficiency.

For persistent storage, satellites differ from terrestrial commercial products as well.
The harsh environment---including launch stresses, high radiation levels, extreme temperature fluctuations, and vacuum conditions---makes hard disk drives impractical for use in space.
Instead, radiation-hardening techniques are applied to solid-state drives, which are by default more resilient to physical stress.
These include radiation-hardened memory controllers, shielding to mitigate radiation effects, and \gls{ecc} to ensure data integrity.
Although this ensures stable operation, it also induces higher production emissions, which are already significant for commercial off-the-shelf SSDs (approximately \qty{160}{\gCOeq\per\giga\byte}~\cite{tannu2023-dirty-ssds}).
In addition, the launch and re-entry emissions must be added on top of that.

\subsubsection*{Network Intensity}
Finally, a computing system also uses communication networks to exchange with other systems.
Here, we have a packet intensity $I_{\vPacket} = \mathit{OI}_{\vPacket} + \mathit{MI}_{\vPacket}$ in \unit{\gCOeq\per\pkt}.
We assume $\mathit{MTU} = \qty{1500}{\byte}$ with $T_{\vTransmission} = \frac{\mathit{MTU}}{\vDataRate}$.
If the transceiver has a power $P_{\vTransceiver}$, we get $E_{\vPacket} = T_{\vTransmission} \cdot P_{\vTransceiver}$.
Using the energy intensity, we get $\mathit{OI}_{\vPacket} = E_{\vPacket} \cdot I_{\vEnergy}$.
On the embodied side, we get the total embodied emissions $\mathit{TE}_{\vTransceiver}$ as usual from the production and launch emissions varying with the launch technology.
In the \gls{sci} mindset, $\mathit{TS} = \frac{T_{\vTransmission}}{T_{\vMission}}$, hence $\mathit{MI}_{\vPacket} = \mathit{TS} \cdot \mathit{TE}_{\vTransceiver}$.

\begin{mintedfloat}[t]
    \begin{minipage}[t]{.48\linewidth}
        \begin{minted}[
            frame=lines,
            fontsize=\small,
            escapeinside=||,
            autogobble,
            linenos=true,
            numbersep=6pt
            ]{toml}
            # sysX.toml (X |$\in$| {E, F, S})

            mission_time = 5.0 # yrs
            launch = "None" # or "Falcon9" or "Starship"|\phantomsection\label{configline}|

            [solar_array]
            area = 2.0 # m^2
            panels = 1 # 1

            [battery]
            capacity = 4.0 # kWh
            cycles = 5000  # 1

            [transceiver]
            mass = 24.5      # g
            power = 1        # W
            data_rate = 38.4 # Kbps
        \end{minted}
    \end{minipage}%
    \begin{minipage}[t]{.52\linewidth}
        \begin{minted}[
            frame=lines, 
            fontsize=\small, 
            autogobble, 
            escapeinside=||
            ]{toml}
            |\vphantom{y}|

            |\vphantom{y}|
            |\vphantom{p}|

            [cpu]
            mass = 100          # g
            area = 1137.5       # mm^2
            max_power = 28.0    # W

            [dram]
            capacity = 16.0          # GB
            power_per_memory = 0.020 # W/GB

            [ssd]
            capacity = 4.0      # TB
            average_power = 3.0 # W
        \end{minted}
    \end{minipage}%
    \caption{
        Configurations for the
            {\textcolor{tol-hc-yellow}{\faIcon{earth-europe} E(arth)}},
        {\textcolor{tol-hc-red}{\faIcon{rocket} F(alcon-9)}},
        and {\textcolor{tol-hc-blue}{\faIcon{rocket} S(tarship)}} systems.
        Line~\ref{configline} specifies the technology.
    }
    \label{lst:sys-conf}
\end{mintedfloat}

\subsection{Estimating}
\begin{figure*}[t]
    \centering%
    \input{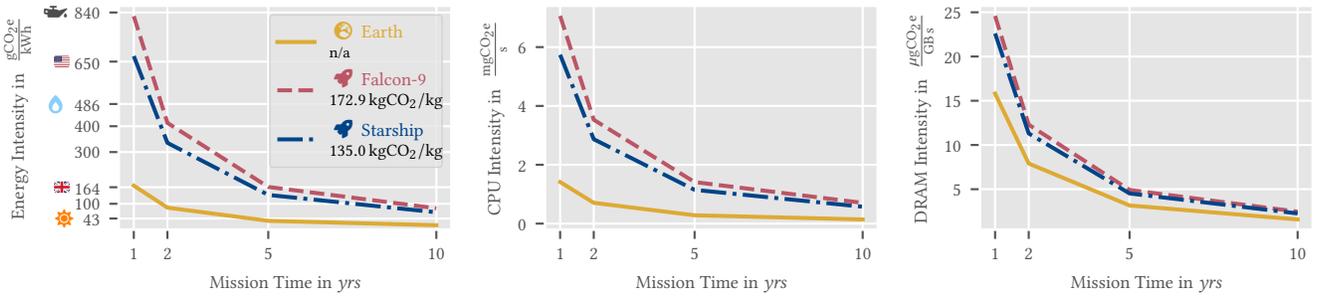}%
    \caption{Increasing the mission time reduces the carbon intensities of operational energy, as well as computation and storage.
        However, keeping things terrestrial or using a less carbon-intensive technology can help to reduce emissions.
        The energy intensities have reference values for terrestrial energy sources (abstract symbol) and 2025 country mixes (national flag).}
    \label{fig:intensities}
\end{figure*}

We implemented the sustainability models in a tool named \sThree{}.
The tool reads one (or more) \texttt{System\allowbreak{}Config.toml} (cf.\@ \autoref{lst:sys-conf}) and applies our mathematical framework.
Thereby, it produces \texttt{System\allowbreak{}Report.md} files~\cite{espas-zenodo} for visual analysis~(\autoref{fig:intensities}).
The tool is written in Rust and leverages the \texttt{uom} crate~\cite{rust:crate-uom} for dimensional analysis and automatic conversions---giving us confidence that the computations are correctly implemented.

\section{Worked Examples}
\label{sec:worked-examples}

In the following, we take a closer look at two aspects of data centers in orbit: how the launch technology impacts these and how a workload's footprint compares between Earth and orbit.

\subsection{Launch Technologies}
We compare three systems, dimensioned for five year missions, that are identical except their choice of launch technology:
\textcolor{tol-hc-yellow}{\faIcon{earth-europe}~Earth}, i.e. no launch, \textcolor{tol-hc-red}{\faIcon{rocket}~F9}, and \textcolor{tol-hc-blue}{\faIcon{rocket}~StSh}.
If we look at the energy intensity of the solar array $I_{\vEnergy, \vSolarArray}$, we see that it halves in orbit~(\qty{14.0}{\gCOeq\per\kWh} to \qty{7.5}{\gCOeq\per\kWh} for F9 or \qty{6.3}{\gCOeq\per\kWh} for StSh).
This is due to the fact that the lack of atmosphere increases the efficiency of the panels significantly~(even outweighing the launch and re-entry cost).
In discussions, this relation is often used as an argument for ``energy is (carbon-) free in space''.
What is often overlooked is that solar arrays do not work at all times, but eclipse times make a battery necessary~(also on Earth, but the mass is neither launched nor re-entered).
Taking this into account, we arrive at $I_{\vEnergy, \vEarth} = \qty{34.0}{\gCOeq\per\kWh}$, $I_{\vEnergy, \vRocket} = \qty{66.3}{\gCOeq\per\kWh}$ and $I_{\vEnergy, \vStarship} = \qty{52.1}{\gCOeq\per\kWh}$.
The left plot of \autoref{fig:intensities} shows mission-time dependent curves of $I_{\vEnergy, t}$ for the different technologies~$t$.
The left axis references terrestrial energy sources~(\textcolor{c-gas}{\faIcon{wind}}~Wind, \textcolor{orange}{\faIcon{sun}}~Solar) as well as 2025 country mixes~(\twemoji{finland}~FI, \twemoji{gb}~GB).
\insightbox{Reducing launch and re-entry intensities has a detectable but low impact on the (energy) intensity.}

\subsection{Data Center Workloads in Orbit}
\label{subsec:workload-in-orbit}

We compare how costly it is to run a comparable data center workload on Earth and in orbit.
The middle and right plots of \autoref{fig:intensities} show the CPU and memory intensities as functions of the mission time.
For energy, we see that irrespective of the launch and re-entry intensity, the shape of the curve is the same: longer missions are better---especially extending short ones is valuable.
Now, looking at the concrete launch technology, we see that it creates a gap between the curves that is persistent.
What is more interesting is that launch and re-entry are significant, but not the only contributors to the overall intensity.
Coming back to a computing workload, we compare \emph{idle} intensities (the hardware is put into place, but shut off, consuming no power, i.e. $M$) with \emph{full} load intensities~(the system runs at maximum power, i.e. $M+O$) (in \autoref{tab:intensities}).
\insightbox{Data center workloads in orbit cause significantly more carbon emissions compared to being computed on Earth.}

\begin{table}
    \caption{Idle ($M$) and Full Load ($O+M$) workload intensities for earthbound operation and different launch technologies.}
    \label{tab:intensities}
    \small%
    \begin{tabular}{lcrrr}
        \toprule
        \shortstack[l]{\textbf{\faIcon{puzzle-piece} Component}                                           \\\footnotesize Intensity Unit} & \textbf{SCI} & \textbf{\textcolor{tol-hc-yellow}{\faIcon{earth-europe} Earth}} & \textbf{\textcolor{tol-hc-red}{\faIcon{rocket} F9}} & \textbf{\textcolor{tol-hc-blue}{\faIcon{rocket} StSh}} \\
        \midrule
        \multirow{2}{*}{\shortstack[l]{\textbf{\faIcon{microchip} CPU}                                    \\
        \footnotesize\unit{\micro\gCOeq\per\second}}}               & $M$   & $18.0$  & $50.8$  & $39.8$  \\
                                                                    & $O+M$ & $282.8$ & $566.3$ & $445.4$ \\
        \midrule
        \multirow{2}{*}{\shortstack[l]{\textbf{\faIcon{memory} DRAM}                                      \\
        \footnotesize\unit{\micro\gCOeq\per\giga\byte\per\second}}} & $M$   & $3.0$   & $3.3$   & $3.2$   \\
                                                                    & $O+M$ & $3.2$   & $3.6$   & $3.5$   \\
        \midrule
        \multirow{2}{*}{\shortstack[l]{\textbf{\faIcon{hard-drive} SSD}                                   \\
        \footnotesize\unit{\micro\gCOeq\per\giga\byte\per\second}}} & $M$   & $0.040$ & $0.045$ & $0.043$ \\
                                                                    & $O+M$ & $0.047$ & $0.058$ & $0.054$ \\
        \midrule
        \multirow{2}{*}{\shortstack[l]{\textbf{\faIcon{satellite-dish} Transceiver}                       \\
        \footnotesize\unit{\micro\gCOeq\per\pkt}}}                  & $M$   & $4.2$   & $6.7$   & $5.9$   \\
                                                                    & $O+M$ & $7.1$   & $12.4$  & $10.4$  \\
        \bottomrule
    \end{tabular}
\end{table}

\subsection{In-Orbit Data Aggregation}
\newcommand{\vData}{\ensuremath{\mathit{Data}}}
\newcommand{\vAggregated}{\ensuremath{\mathit{Aggr}}}
\newcommand{\vHop}{\ensuremath{\mathit{Hop}}}
\newcommand{\vHops}{\ensuremath{\mathit{Hops}}}
\newcommand{\vISL}{\ensuremath{\mathit{ISL}}}
\newcommand{\vGSL}{\ensuremath{\mathit{GSL}}}
\newcommand{\vProcessing}{\ensuremath{\mathit{Proc}}}
\newcommand{\vNetworking}{\ensuremath{\mathit{Netw}}}

\begin{figure}[t]
    \centering%
    \input{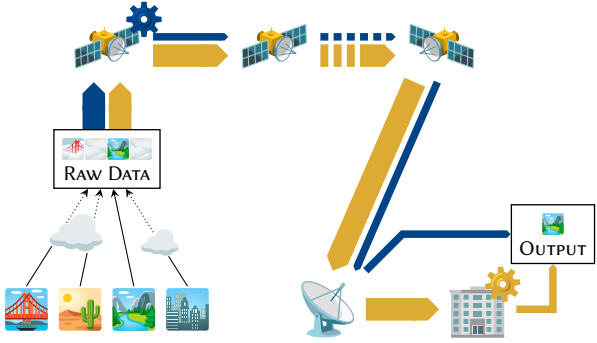}%
    \caption{Comparison of processing data in orbit (\textcolor{tol-hc-blue}{blue path}) or on Earth (\textcolor{tol-hc-yellow}{yellow path}). The arrow thickness visualizes the data volume.}
    \label{fig:processing-space-ground}
\end{figure}

In-orbit computing is often used to aggregate the raw data before downloading it to Earth---effectively reducing the data volume.
We see this in \autoref{fig:processing-space-ground}, where the thickness of arrows visualizes the volume of traffic.
An example of this is ground observation, where cloudy images can be discarded early.
Note that, following the idea of a LEO Internet constellation, the satellite with processing is not necessarily the one with \gls{gsl} capabilities (e.g.\@ because there is (currently) no visibility to a ground station), requiring the forwarding of the data via \gls{isl}~\cite{DBLP:journals/cn/FraireHSOHWBRMSR24}.

Hence, we can formalize the carbon intensity of such a data flow with a post-aggregation rate of $f_{\vAggregated}$ and when using launch technology (or Earth)~$t$:
\begin{align*}
    I_{\mathit{Flow}, t}\big(f_{\vAggregated}\big) & = I_{\vProcessing, t} + I_{\vNetworking, t}\big(f_{\vAggregated}\big)      \\
    I_{\vNetworking, t}\big(f_{\vAggregated}\big)  & = f_{\vAggregated} \cdot \big(I_{\vGSL} + n_{\vHops} \cdot I_{\vISL} \big)
\end{align*}

\begin{figure*}[t]
    \centering%
    \input{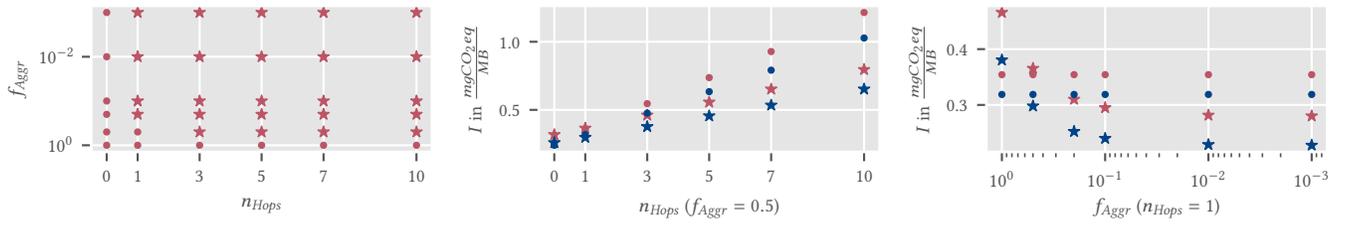}%
    \caption{
        The number of hops and aggregation potential determine carbon intensity.
        We compare computing on Ground~($\bullet$) or in Orbit~($\star$) and consider launch technologies \textcolor{tol-hc-red}{\faIcon{rocket}~F9} and \textcolor{tol-hc-blue}{\faIcon{rocket}~StSh}.
        The left chart shows for which parameter pair which approach is less carbon-intensive (for \textcolor{tol-hc-red}{\faIcon{rocket}~F9}).
        The other charts fix one parameter and show how the absolute carbon intensity changes depending on the other parameter---the launch technology determines the inflection point.
    }
    \label{fig:inorbit}
\end{figure*}

Here, we use $n_{\vHops}$ with $I_{\vISL}$ and one with $I_{\vGSL}$ for the networking footprint.
We use the $I_{\vProcessing, t}$ as the terrestrial or orbital computing's intensity.
Purely analytically, we see that the concrete answer to ``is in-orbit computing more sustainable?'' depends on the concrete workload's computational effort and its proportion to its raw size, as well as of its size after aggregation $f_{\vAggregated}$.

To shed light on this, we explore the parameter space.
Of central importance is $f_{\vAggregated}$ (as it quantifies the maximum achievable gain in networking footprint) but also the number of hops.
We consider rates $f_{\vAggregated} \in \left\lbrace 100, 70, 50, 20, 10 \right\rbrace\%$ where $100\%$ represents no data reduction.
We also consider $n_{\vHops} \in \left\lbrace 0, 1, 2, 3\right\rbrace$, where 0 means that the raw data is produced (and potentially) processed at the satellite immediately connected to the ground station.

\pagebreak 

In \autoref{fig:inorbit} (left) we use one system launched via \textcolor{tol-hc-red}{\faIcon{rocket}~F9} and mark which combination of  $f_{\vAggregated}$ and  $n_{\vHops}$ causes less carbon in \textbf{$\star$~Orbit} or on \textbf{$\bullet$~Ground}.
The resulting frontier suggests that only with a high $f_{\vAggregated}$ and a small $n_{\vHops}$, computing on ground is superior to in-orbit computing.
Going into details in \autoref{fig:inorbit} (middle), where we have $f_{\vAggregated} = 0.7$, in-orbit compute makes sense as soon as \glspl{isl} must be used---the inflection points are $n_{\vHops} = 1$ (F9) or $2$ (StSh).
In \autoref{fig:inorbit} (right), with $n_{\vHops} = 1$, in-orbit computing is better as soon as the data can be aggregated down to 50\%~(F9) or 70\%~(StSh).
We further see that the choice of launch technology shifts the inflection points slightly, but does not change the general trend.

\insightbox{
    The sustainability of in-orbit computing depends heavily on the characteristics of the application---especially the relation between raw data size, processing footprint, and potential aggregation rate.
}

\section{Conclusion}
\label{sec:conclusion}

At first glance, putting computation in \gls{leo} sounds attractive: thanks to single-AS routing, the latency could outperform what many terrestrial routes offer.
The location of the satellites makes for good global availability.
Further, reusable launch technology creates hope for improved sustainability in the future.
At second glance, however, in particular with a sustainability lens, computing in \gls{leo} is costly for the environment.
An obvious contributor is the launch---which offsets positive aspects of space operations~(e.g. increased solar panel efficiency due to missing atmosphere).
A second, often overlooked, contributor is re-entry, where payload and rocket pieces burn, turning into \unit{\NOx} that is significantly more dangerous than~\(\mathrm{CO_{2}}\).

This paper has shown that even with improving the launch technology, orbital applications will not beat terrestrial ones.
For the future, we see the following avenues to improve sustainability.
Despite the relatively low impact mentioned above, improving launch technology can still reduce the emission intensity (\unit{\kgCOeq} per \unit{\kilo\gram} payload).
Reducing the mass of components (the \unit{\kilo\gram}) is another option.
Finally, given the positive effect of a longer mission time, striving for this goal is worthwhile.
Additionally, studies should extend upon \citeauthor{osoro2023sustainability}~\cite{osoro2023sustainability} and not just compare residential access footprints, but compare how a terrestrial and a \gls{leo} backbone network compare in a more detailed manner.
Whether the new space race will be one for sustainability, we will have to see.

\section*{Availability}

\sThree{} is dual-licensed under the Apache-2.0 or MIT license.
The latest source code is publicly available at \url{https://gitlab.com/sustainable-computing-systems/espas}, and is also archived on Zenodo~\cite{espas-zenodo}.

\begin{acks}
    This project has received funding from the European Union's Horizon 2020 research and innovation programme under the Marie Skłodowska-Curie grant agreement \href{https://doi.org/10.3030/101008233}{No 101008233} -- MISSION, see \url{https://mission-project.eu}, and by DFG grant 389792660 as part of TRR~248 -- CPEC, see \url{https://perspicuous-computing.science}.
\end{acks}

\patchcmd{\bibsetup}{\interlinepenalty=5000}{\interlinepenalty=10000}{}{}
\printbibliography

\end{document}